# Graph Neural Network Reveals the Cortical Morphology of Local Brain Aging in Normal Cognition and Alzheimer's Disease

Samuel D. Anderson[1], Jordan Jomsky[1,2], Nikhil N. Chaudhari[1,2], Nahian F. Chowdhury[1,3], Xiaoyu (Rayne) Zheng[4,5], and Andrei Irimia[1,2,3,6,7], for the Alzheimer's Disease Neuroimaging Initiative

1 Ethel Percy Andrus Gerontology Center, Leonard Davis School of Gerontology, University of Southern California, Los Angeles, CA 90089, USA

2 Corwin D. Denney Research Center, Alfred E. Mann Department of Biomedical Engineering, Viterbi School of Engineering, University of Southern California, Los Angeles, CA 90089, USA

3 Neuroscience Graduate Program, University of Southern California, Los Angeles, CA 90089, USA

4 Department of Materials Science and Engineering, University of California, Berkeley, Berkeley, CA 94720, USA

5 Lawrence Berkeley National Laboratory, Berkeley, CA 94720, USA

6 Department of Quantitative & Computational Biology, Dana & David Dornsife College of Arts & Sciences, University of Southern California, Los Angeles, CA, USA

7 Centre for Healthy Brain Aging, Institute of Psychiatry, Psychology & Neuroscience, King's College London, UK

**Correspondence should be addressed to:**

Andrei Irimia (irimia@usc.edu)

**Abstract**

Estimating brain age (BA) from $T_1$-weighted magnetic resonance images (MRIs) provides a powerful framework for quantifying anatomical brain aging. Whereas *global* BA (GBA) summarizes overall brain health, *local* BA (LBA) provides cortically specific patterns of aging at the subject level. Although previous studies have examined anatomical contributors to GBA, to our knowledge, no framework has been established to estimate LBA using cortical morphology. To address this gap, we introduce a graph neural network (GNN) that uses morphometric features—cortical thickness, surface area, curvature, gray/white matter intensity ratio (GWR), sulcal depth—to estimate LBA across the cortical surface at high spatial resolution (mean inter-vertex distance = 1.37 mm). Trained on cortical surface meshes extracted from the MRIs of cognitively normal (CN) adults ($N$ = 14,423), our model achieves lower mean absolute error (MAE) than the existing state-of-the-art while identifying more biologically plausible patterns of aging in Alzheimer's disease (AD) on the ADNI dataset. Association cortices emerge as primary sites of morphometric aging in CNs, whereas mild cognitive impairment is characterized by widespread aging that is pronounced in the parahippocampal gyrus. AD subjects demonstrate significant aging across the entire cortex, particularly within medial temporal regions and associated cortical networks. Feature ablation highlights curvature and GWR as preferentially sensitive to AD pathology. Regional LBA gaps are significantly associated with neuropsychological measures of AD-related cognitive impairment, linking cortical aging patterns to clinical outcomes. These results demonstrate that GNN-based modeling of cortical morphometry enables biologically interpretable mapping of local brain aging with greater interpretability than prior work.

## Introduction

The cortical surface undergoes significant changes during normative aging and plays a central role in the pathogenesis of neurodegenerative disorders such as Alzheimer's disease (AD) [1-3]. Changes in cortical thickness [4, 5], surface area [6, 7], sulcal depth [8, 9], gray/white matter intensity ratio (GWR) [10, 11], and curvature [12, 13] are all hallmarks of brain aging. Deviations from normative patterns of cortical aging manifest early in AD and may even precede clinical symptoms [1, 2, 14]. As such, there is value in studying how cortical features vary with age in both cognitively normal (CN) and cognitively impaired (CI) populations.

Although the effects of chronological age (CA) on cortical aging have been studied extensively, CA alone does not adequately capture the heterogeneity of aging effects across individuals. This has motivated the development of tools capable of quantifying the biological age of the brain, i.e., brain age (BA), grounding measures of disease risk on the observed variances of aging profiles [15-17]. The difference between BA and CA—termed BA gap (BAG)—reflects advanced (BAG > 0 y) or delayed (BAG < 0 y) brain aging, and may serve as a candidate biomarker of excessive aging observed in neurodegenerative disorders such as AD [18-20].

Typically, BA is computed as a single global value—global BA (GBA)—which summarizes aging across the entire brain [18, 21, 22] or cortex [23-25]. While informative, this approach requires regional differences in brain aging to be localized *indirectly*, often via saliency methods which may be unreliable in biomedical contexts [26-29]. In contrast to GBA, local BA (LBA) computes vertex- or voxel-level BA, yielding local BAGs (LBAGs) that *directly* capture spatial variability in aging across the brain. An LBA model generates a spatial map of the brain by utilizing measured attributes—such as local cortical thickness or magnetic resonance image (MRI) intensity—to estimate LBA at each vertex or voxel of the cortical surface. In contrast to attention or pooling-based approaches [24, 25], which cannot directly associate subject-level local aging with outcome variables [30, 31], these maps can be used to relate local aging patterns to phenotypic measures [32], thereby enhancing the utility and specificity of BA modeling in both research and clinical contexts [33, 34].

Previous LBA models have been hindered by both feature selection and methodology. With respect to feature selection, existing methods leverage MRI intensities derived from $T_1$-weighted ($T_1$w) MRI to generate LBA estimates [35-37]. While these intensities encode subtle molecular and structural interactions within brain tissue [38], their relationship to cortical morphology is complex [39], limiting the biological interpretability of provided LBAs. Methodologically, prior approaches have relied upon convolutional neural networks (CNNs) to extract aging patterns. CNNs have achieved great success across diverse tasks, but impose several key assumptions which may limit their applicability to LBA estimation. Namely, CNNs draw inferences by dividing the brain into volumetric cubes of arbitrary sizes [40], causing anatomically distant regions to be treated as topologically close. Additionally, CNNs assume neighborhood regularity, which conflicts with the highly irregular geometry of the cortex [41-44]. Cumulatively, these factors may cause CNNs to conflate separate patterns of local aging, making them sub-optimal for the estimation of LBA.

To our knowledge, we introduce the first graph neural network (GNN)-based approach to LBA. Our approach conceptualizes the cortical surface, derived from $T_1$w MRI, as a three-dimensional surface mesh, with morphometric features mapped to each vertex along this mesh. This mesh and its features are used as input to the model, facilitating the use of explainability techniques—such as feature ablation—to identify the morphological underpinnings of local aging. By preserving the geometric and topological continuity of the cortical surface, our approach enables anatomically faithful modeling of inter-regional dependencies. Recent work motivates this approach: graph-based methods achieve state-of-the-art performance in surface-based GBA computation [24], and outperform alternative methods—such as CNNs or vision transformers—in image segmentation [45-47], where accurate delineation of fine anatomical boundaries is critical. Collectively, these properties facilitate the robust quantification of cortical aging patterns, providing a scalable foundation for studying individual variability in morphometric aging across CN and CI populations alike.

## Methods

**Data.** $T_1$w MRI scans were aggregated across multiple sources to enhance the generalizability of our model and findings. All participants provided written informed consent at their respective contributing institutions. This study was conducted in accordance with the U.S. Code of Federal Regulations (45 C.F.R. 46) and the Declaration of Helsinki. The training sample was comprised of CN adults from the UK Biobank (UKBB), National Alzheimer's Coordinating Center (NACC), and Information eXtraction from Images (IXI) datasets [48-50], totaling 14,423 scans spanning a broad CA range. Model evaluation was performed using the Alzheimer's Disease Neuroimaging Initiative (ADNI) dataset, which included 1,129 scans from CN participants, 153 scans from MCI subjects, and 477 scans from individuals with a clinical diagnosis of AD [51]. Participant demographics and dataset characteristics are summarized in **Table 1**.

| | | | Chronological Age | | | | | |
|---|---|---|---|---|---|---|---|---|
| Repository | Set | $N$ | Min | Max | $\mu$ | $\sigma$ | M:F | FS version |
| UKBB | training | 9,619 | 45.5 | 82.4 | 64.8 | 7.8 | 1:1.1 | 6.0.0 |
| NACC | training | 4,324 | 18.9 | 100.2 | 69.2 | 10.9 | 1:2.0 | 7.1.1 |
| IXI | training | 480 | 20.0 | 86.3 | 50.9 | 16.1 | 1:1.3 | 6.0.0 |
| All | training | 14,423 | 18.9 | 100.2 | 65.7 | 9.8 | 1:1.3 | – |
| ADNI (CN) | testing | 1,129 | 55.5 | 104.3 | 75.7 | 6.8 | 1:1.0 | 6.0.0 |
| ADNI (MCI) | testing | 153 | 56.5 | 95.9 | 73.7 | 7.7 | 1:1.2 | 6.0.0 |
| ADNI (AD) | testing | 477 | 55.2 | 93.0 | 76.1 | 8.1 | 1:0.9 | 6.0.0 |

**Table 1. Dataset statistics.** Number of scans ($N$), descriptive statistics of chronological age (CA; minimum Min, maximum Max, mean $\mu$, standard deviation $\sigma$), the male-to-female (M:F) ratio, and FreeSurfer version (FS version) used for MRI preprocessing.

The ADNI was launched in 2003 as a public–private partnership, led by Principal Investigator Michael W. Weiner, MD. The primary goal of ADNI has been to test whether serial MRI, positron emission tomography, other biological markers, and clinical and neuropsychological assessment can be combined to measure the progression of mild cognitive impairment (MCI) and early AD. For ADNI CN adults, inclusion criteria are: no memory complaints, a Clinical Dementia Rating (CDR) of zero, no significant impairment in cognitive function or activities of daily living, and a score of at least nine on the Logical Memory II subscale of the Wechsler Memory Scale–Revised. For NACC, participants were included in the training set only if physicians deemed them CN based on cognitive assessment and/or personal and medical history. Some participants contributed scans acquired at different time points. In the training sample, all UKBB and IXI participants contributed only a single scan, whereas NACC contributed 4151 scans across 3047 participants. These scans were retained as independent samples during training to mitigate the over-representation of UKBB [52-55]. In the testing sample, the ADNI CN cohort comprised 1,129 scans from 517 unique participants, while the ADNI MCI and AD cohorts comprised 153 and 477 scans from 83 and 354 unique participants respectively.

**Preprocessing.** $T_1$w MRI scans were preprocessed using FreeSurfer (FS) to extract high-resolution cortical surface features. These features served as inputs to our GNN model and included cortical thickness, sulcal depth, curvature, surface area, and GWR. The FS pipeline included intensity normalization, skull stripping, bias field correction, and segmentation of brain tissue into white matter (WM), gray matter (GM), and cerebrospinal fluid (CSF). FS reconstructs cortical surfaces by initializing a tessellated sphere within the WM, and iteratively deforming it outward towards the GM–CSF boundary, guided by local intensity gradients. This process ensures anatomically precise localization of cortical structures and the extraction of vertex-level features used in our model [56]. Importantly, this procedure yields a single self-contained mesh per sample and hemisphere. To facilitate sample-level learning, we project each mesh onto a standardized, hemisphere-specific cortical atlas [57], then concatenate these atlas-projected meshes to derive a single, whole-cortex mesh per sample. In our GNN model, mesh vertices function as graph nodes, while edges connect cortically adjacent vertices within each hemisphere [23, 24, 42]. Because all cortical surfaces are resampled onto the same standardized atlases, these sample-level graphs differ only in the morphometric features of their nodes.

**Atlas resampling.** To enable learning across multiple feature scales while preserving anatomical structure as an inductive bias, we developed a pooling/unpooling strategy that projects learned feature representations across FS atlas resolutions. To this end, we defined *receptive fields* for each vertex, at each resolution, in a bottom-up manner.

First, we derived Euclidean coordinates for each vertex in each resolution using spherical projections of each FS atlas. For each vertex $v$ in resolution $i$, termed $v_i$, we used these coordinates to identify the closest vertex in resolution $i+1$, allowing us to identify each vertices' equivalent in the higher resolution mesh, termed $v_{i+1}$. We then identified which vertices in resolution $i+1$ were connected to $v_{i+1}$, i.e., the 1-hop spatial adjacency of $v_{i+1}$. The matched vertex $v_{i+1}$ together with its 1-hop neighbors define the receptive field of $v_i$, denoted as $R(v_i)$. See **Figure 1** for a visual representation.

We further define the inverse of the receptive field such that, for any vertex $v_{i+1}$, its inverse receptive field $R^{-1}(v_{i+1})$ is the set of all lower-resolution vertices $v_i$ whose receptive field contains $v_{i+1}$. The inverse receptive field of a vertex $v_{i+1}$ can therefore be understood as the collection of all vertices $v_i$ which have $v_{i+1}$ in *their own* receptive fields:

$$R^{-1}(v_{\mathrm{i+1}}) = \{v_{\mathrm{i}} \mid v_{\mathrm{i+1}} \in R(v_{\mathrm{i}})\}$$

Downsampling is performed by averaging features within each receptive field, producing one representation per lower-resolution vertex. Conversely, upsampling is achieved by averaging across inverse receptive fields. Both operations yield feature matrices aligned to their respective resolutions, enabling smooth information flow across scales.

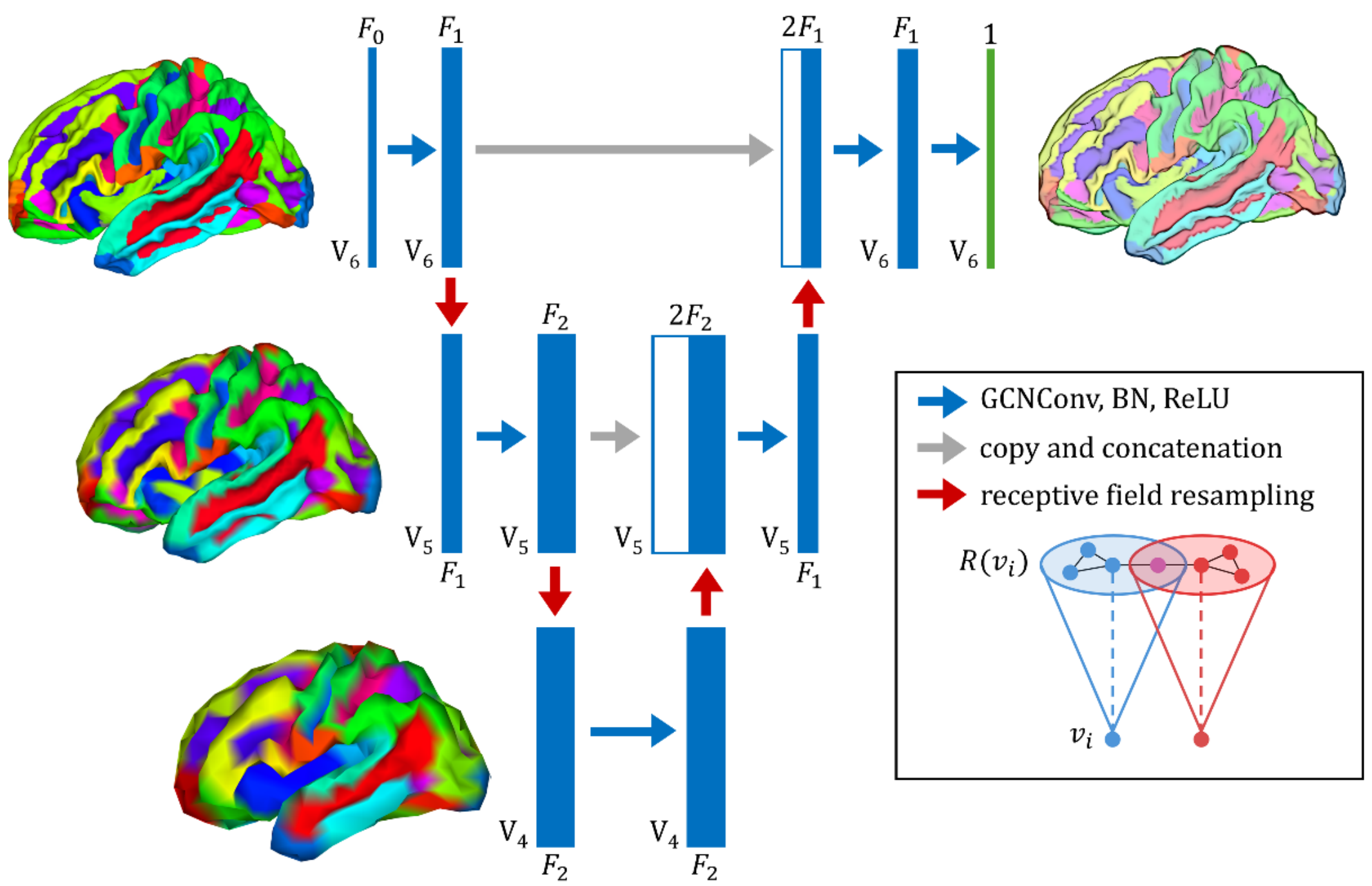

**Figure 1. Graph U-Net architecture for cortical surface modeling.** Blue rectangles denote feature maps defined on the cortical surface mesh at different atlas resolutions (e.g., $V_4$, $V_5$, $V_6$). Joined blue and white rectangles represent skip connections, where features from earlier blocks are concatenated with those from later blocks. Each rectangle is annotated with the number of features $F$ and vertices $V$. The final output (green box with cortical rendering) is a vertex-wise cortical map predicting age at each surface location. Note that the last convolution does not undergo BN or ReLU. The supplementary box provides a visualization of the receptive fields used for resampling. The dashed line associates $v_i$ with its coordinate-matched vertex $v_{i+1}$, the circles represent receptive fields, and the colors represent inverse receptive fields. Averaging across receptive fields lowers resolution, while averaging across inverse receptive fields increases resolution.

**Model structure**. As shown in **Figure 1**, the model follows a hierarchical U-Net [58] architecture designed for hemisphere-concatenated cortical meshes in atlas space. It consists of five blocks arranged across three FS atlas resolutions: ico6 ($V_6$ = 81,924 vertices), ico5 ($V_5$ = 20,484 vertices), and ico4 ($V_4$ = 5,124 vertices), following the sequence ico6 → ico5 → ico4 → ico5 → ico6. The first two blocks form the encoder and the last two form the decoder. Within each block, feature representations are derived or refined using a single graph convolutional network layer (GCNConv) [59] followed by batch normalization (BN) [60] and a rectified linear unit (ReLU) activation function [61]. The input block receives five morphometric features per vertex ($F_0$ = 5), while the number of derived feature channels in subsequent blocks ($F_1$, $F_2$) is treated as a tunable hyperparameter.

In the encoder, the cortical mesh is progressively downsampled from ico6 to ico4 via receptive field resampling layers between consecutive blocks. In the decoder, this process is reversed: receptive field resampling layers between blocks restore higher-resolution feature representations from ico4 back to ico6. Skip connections [62] link encoder and decoder blocks at matching resolutions, preserving fine-grained spatial information throughout the network. In the final decoder block, an additional GCNConv layer is applied after BN and ReLU. The network outputs a single value per vertex representing the estimated LBA at ico6.

**Model training.** Model hyperparameters—including feature sizes, batch size, learning rate, weight decay, and number of epochs—were optimized using an 80/20 subject-level training/validation split derived from the training data. Hyperparameters were selected via grid search based on validation loss. After optimization, the training and validation subsets were recombined into a single training set for final model fitting. Model weights were initialized using the default PyTorch Geometric initialization scheme (Kaiming uniform initialization [63]). All analyses were performed on a single NVIDIA A100 GPU. The model was optimized using the AdamW optimizer [64] with feature sizes of $F_1$ = 256 and $F_2$ = 512, a batch size of 8, a learning rate of 0.0005, and a weight decay of 0.0003 for 60 epochs. Vertex-wise mean absolute error (MAE) was used to compute loss. Model performance was evaluated using the average MAE across all vertices for ADNI CN, which served as an independent test set to assess generalizability to unseen data.

**Medial wall removal, clipping, and smoothing**. After the model has completed training and generated predictions for the test cohorts, we remove the medial wall from further consideration as it is not a part of the cortex. We then clip LBA predictions from the 1st to 99th percentiles to remove large outliers [65], before applying smoothing to the remaining predictions by averaging each node's LBA value with those of its neighboring nodes [66]. When smoothing, we included neighbors up to two steps away and repeated the averaging process four times, removing visually apparent model artifacts while retaining cortical variability (**Figure S1**). When evaluating on the ADNI CN cohort, MAE was computed after clipping and smoothing but prior to bias correction.

**Bias correction**. Previous studies have demonstrated that BA estimates exhibit systematic bias toward the mean of model training data, with younger individuals consistently overestimated and older individuals underestimated [67]. To address this, we performed bias correction. At each vertex, we regressed LBAG on CA [67, 68] to produce a unique slope and intercept per vertex. These coefficients were then averaged across all vertices to obtain a single slope and intercept for the entire cortex. We used these cortex-level coefficients to adjust each subject's LBA based on their CA, retaining potentially biologically relevant local variation while correcting for global biases in model predictions [37].

Formally, let $LBA_{v,s}$ denote the computed LBA for vertex $v$ and subject $s$, and $CA_s$ the CA for subject $s$. LBAG is defined as $LBAG_{v,s} = LBA_{v,s} - CA_s$ and serves as the dependent variable. For each vertex, we regressed $LBAG_{v,s}$ on $CA_s$ across subjects, obtaining a slope $m_v$ and intercept $b_v$:

$$LBAG_{v,s} = m_v CA_s + b_v$$

This yields $V_6 = 81{,}924$ vertex-specific slopes $m_v$ and intercepts $b_v$. We then averaged $m_v$ and $b_v$ across all $v$ to obtain slope $m_\mu$ and intercept $b_\mu$. The averaged coefficients yield an adjustment term $m_\mu CA_s + b_\mu$ which we apply to every $v$ for a given $s$. The corrected LBA per vertex and subject ${LBA^c}_{v,s}$ is thus defined as:

$${LBA^c}_{v,s} = LBA_{v,s} - (m_\mu CA_s + b_\mu)$$

In our analysis, we first bias corrected the ADNI CN cohort, yielding $m_\mu^{CN}$ and $b_\mu^{CN}$. These coefficients were used for all subsequent analyses to ensure that comparisons across cohorts were made with respect to a consistent estimate of sample and model bias [69]. Notably, model losses are always reported *before* bias correction [70].

**Comparing across cohorts.** When comparing across cohorts, we computed differences in bias-corrected LBAGs to quantify group-specific aging effects, yielding ΔLBAGs. These were calculated by subtracting the LBAGs of the reference cohort from those of the cohort of interest. For AD, we computed AD − CN; for MCI, MCI − CN; and for sex, Female − Male.

**Feature ablation.** To assess the contribution of each morphometric feature to model predictions, we performed feature ablation on both the ADNI CN and AD cohorts. Baseline predictions were first obtained by computing bias-corrected LBAs for each cohort $c$, yielding $LBA^c$. We then repeated inference multiple times. Upon each iteration of inference, we set a single feature $f$ to zero, corresponding to replacing subject-level data with the population mean of the training distribution. This procedure yielded a set of ablated predictions $LBA_f^c$, representing model outputs with feature $f$ ablated (e.g., predictions made using CNs with ablated cortical thickness). For each feature and cohort, we next computed the ablation effect α as $\alpha_f^c = LBA^c - LBA_f^c$, quantifying how model predictions change when variability in feature $f$ is removed. Then, to isolate disease-specific effects, we compared ablation effects across cohorts by computing $\Delta\alpha_f = \alpha_f^{AD} - \alpha_f^{CN}$. This contrast measures the extent to which a feature contributes differently to the model's predictions in ADs relative to CNs. A positive $\Delta\alpha_f$ indicates that the feature contributes more strongly to disease-specific aging, while a negative $\Delta\alpha_f$ indicates that the feature contributes more strongly to normal aging. Values near zero indicate that the feature affects both groups similarly.

**Regressing cognitive scores.** To test whether higher LBAGs were associated with poorer cognitive performance, we regressed LBAGs against cognitive scores for each cohort. MCI subjects were excluded due to limited availability of cognitive test data. Trail Making Test B (TMT-B) scores equal to 300 were excluded to remove ceiling effects [71]. The specific cognitive tests analyzed were chosen based on prior BA work [20]. All scores were standardized and transformed so that higher values consistently indicated worse performance, making regression coefficients directly comparable across tests. We performed separate univariate linear regressions, with BAG as the dependent variable and the normalized cognitive score as the predictor of interest, while controlling for CA, sex, and years of education. All *p*-values were corrected for multiple comparisons using the Benjamini–Hochberg procedure, grouped by cognitive test.

**Comparison to LBA-SotA.** To benchmark our model against the existing state-of-the-art CNN model for LBA (LBA-SotA), we compared both MAE and local sensitivity. MAE was compared across all spatial units: the LBA-SotA reports MAE averaged across voxels, whereas our model reports MAE averaged across cortical vertices. Both models were trained on similar cohorts. Evaluation using identical input features was not feasible due to fundamental differences in representation. The LBA-SotA operates on volumetric intensity data in native space, whereas our model uses atlas-aligned representations with derived features defined over a cortical mesh. Thus, local comparisons were performed at the level of matched scans, while MAE comparisons were based on reported ADNI CN performance metrics. Accordingly, these comparisons are intended to evaluate relative predictive performance and spatial sensitivity rather than establish identical architectural benchmarking conditions. Mean LBAGs were computed for each lobe based on each model's LBAGs for ADNI CN and AD subjects. Uncertainty was estimated using the standard error of the mean. To compare spatial patterns independent of global scale, lobe-wise differences were normalized within each model such that their absolute values summed to 100%, yielding the relative contribution of each lobe to pathological aging. This framework enables comparison of where each model attributes AD-related effects and how these patterns align with existing literature.

**Sex comparisons.** To assess the model's responsiveness to sex-specific morphometry, we computed ΔLBAGs between female and male subjects within the ADNI CN cohort. To account for differences in CA distributions, we implemented an age-stratified subsampling procedure. Subjects were grouped into 1-year CA bins, and within each bin, equal numbers of males and females were randomly selected without replacement. This yielded matched subsets with comparable CA distributions, facilitating local comparison.

## Results

**Model Performance.** We found that feature sizes of $F_1$ = 256 and $F_2$ = 512, a batch size of 8, a learning rate of 0.0005, and a weight decay of 0.0003 for 60 epochs yielded optimal model performance on the validation set. The model was fully trained ($N$ = 14,423) using these parameters, and evaluated on the external, independent, ADNI dataset, yielding an average MAE of 5.63 y for the CN cohort prior to bias correction [70]. Our MAE is lower than the LBA-SotA (MAE = 5.94 y) [37], which was trained on similar data and evaluated on ADNI CNs.

**Comparing LBAGs across cohorts.** To determine how local aging patterns differ across cohorts, we examined the bilateral distribution of LBAGs and their cross-cohort differences ΔLBAGs. In CNs, the largest positive effects were observed in visual, parietal, and prefrontal association cortices, including the superior occipital gyrus (2.14 y), angular gyrus (1.02 y), and pars opercularis (0.94 y), with negative LBAGs in the inferior temporal gyrus (-1.96 y) and parahippocampal gyrus (-1.67 y). Weak effects were found in the default mode network (DMN; -0.38 y) and somatomotor network (SMN; 0.29 y), with large negative effects found in limbic/paralimbic regions (LIM; -1.26 y) (**Figure 2A, Table 2**).

Individuals with MCI exhibited widespread positive ΔLBAGs, with relatively uniform effect sizes across regions and networks. The largest ΔLBAGs were found in the pericallosal sulcus (0.80 y) and parahippocampal gyrus (0.79 y), with notably weaker effects in the superior occipital gyrus (0.65 y), pars opercularis (0.54 y), angular gyrus (0.53 y), and inferior temporal gyrus (0.52 y). Network effects were strongest in the LIM (0.64 y) and SMN (0.62 y), and weakest in the DMN (0.49 y) (**Figure 2B, Table 2**).

In ADs, the largest regional ΔLBAGs were observed in the inferior temporal gyrus (2.36 y), with the parahippocampal gyrus (2.04 y) also demonstrating notable increases. ΔLBAGs in the angular gyrus (0.89 y), pars opercularis (0.82 y), and superior occipital gyrus (0.45 y) were relatively weaker. Network ΔLBAGs differed vastly, being strongest in the LIM (1.79 y) and DMN (1.36 y), and weakest in the SMN (0.80 y) (**Figure 2C, Table 2**). All LBAGs and ΔLBAGs are reported by region and network in **Tables S1-2**.

| | Cohort | | |
|---|---|---|---|
| | ADNI CN | ADNI MCI-CN | ADNI AD-CN |
| **Selected Regions** | LBAGs | ΔLBAGs | ΔLBAGs |
| Inferior Temporal Gyrus | -1.96 | 0.52 | 2.36 |
| Parahippocampal Gyrus | -1.67 | 0.79 | 2.04 |
| Superior Occipital Gyrus | 2.14 | 0.65 | 0.45 |
| Angular Gyrus | 1.02 | 0.53 | 0.89 |
| Pars Opercularis | 0.94 | 0.54 | 0.82 |
| **Selected Networks** | | | |
| Default Mode Network | -0.38 | 0.49 | 1.36 |
| Limbic/Paralimbic System | -1.26 | 0.64 | 1.79 |
| Somatomotor Network | 0.29 | 0.62 | 0.80 |

**Table 2. Selected LBAG and ΔLBAG statistics in years.** Representative LBAGs in CN individuals and cross-cohort ΔLBAGs in MCIs and ADs relative to CN participants. Values are shown in years and highlight selected cortical regions and large-scale functional networks demonstrating prominent positive or negative effects across cohorts. Negative LBAGs indicate regions estimated to appear younger than CA, whereas positive values indicate accelerated LBA. ΔLBAGs represent the relative increase in LBAG compared with CN participants. Regions and networks were selected based on within-cohort outlier status and biological priors.

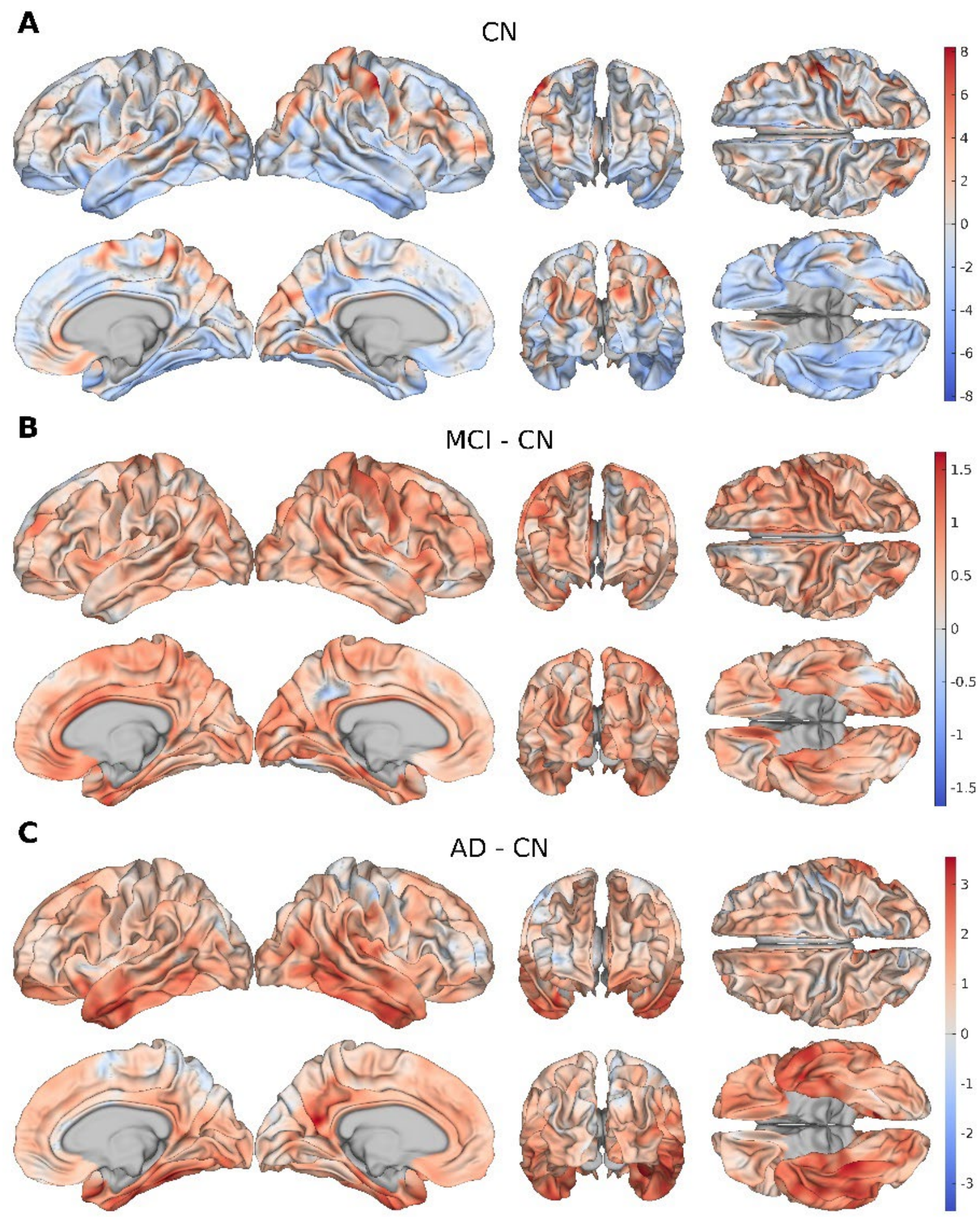

**Figure 2. Bias-corrected LBAGs and ΔLBAGs for ADNI CNs, MCIs, and ADs.** (A) Bias-corrected LBAGs for CN subjects. (B) Vertex-wise ΔLBAGs between MCIs and CNs (MCI – CN). (C) Vertex-wise ΔLBAGs between ADs and CNs (AD – CN).

**Feature ablation.** To quantify the contribution of each morphometric feature to local aging, we performed feature ablation across the CN and AD cohorts, and identified their differences, i.e., $\Delta\alpha_f$. Surface area demonstrated its strongest negative effects throughout the visual and parietal association cortices, with the cuneus (-1.19 y), angular gyrus (-1.03 y), and superior occipital gyrus (-1.00 y) exhibiting the three largest effects. Weaker effects were observed in the parahippocampal gyrus (-0.69 y), pars opercularis (-0.63 y), and inferior temporal gyrus (-0.43 y). No notable divergence was found across networks (SMN = -0.54 y; DMN = -0.53 y; LIM = -0.50 y) (**Figure 3A, Table 3**).

Cortical thickness had its largest $\Delta\alpha_f$ values in the inferior temporal gyrus (1.26 y), with a strong effect in the parahippocampal gyrus (1.00 y) and weaker effects in the angular gyrus (0.28 y), pars opercularis (0.21 y), and superior occipital gyrus (0.17 y). At the network level, the LIM (0.90 y) demonstrated the largest effect, with comparatively weaker effects in the DMN (0.54 y) and SMN (0.38 y) (**Figure 3B, Table 3**).

Similarly, GWR had its largest regional effects in the inferior temporal gyrus (1.98 y), a strong effect in the parahippocampal gyrus (1.77 y), and weaker effects in the angular gyrus (1.15 y), pars opercularis (0.81 y), and superior occipital gyrus (0.45 y). Both the LIM (1.52 y) and DMN (1.32 y) evidenced notable increases, with a lesser effect in the SMN (1.17 y), contrasting with cortical thickness (**Figure 3C, Table 3**).

Curvature followed a similar pattern, with the strongest effects in the inferior temporal gyrus (2.19 y), strong effects in the parahippocampal gyrus (1.54 y), and weaker effects in the pars opercularis (1.29 y), angular gyrus (1.13 y), and superior occipital gyrus (0.54 y). A network preference for the LIM (1.71 y) and DMN (1.53 y) over the SMN (0.94 y) was also observed (**Figure 3D, Table 3**).

Sulcal depth exhibited only weak $\alpha_f^c$ values across all regions and networks (**Figure 3E, Table 3**). All $\alpha_f^c$ and $\Delta\alpha_f$ values are reported by region and network in **Tables S3-4**.

| Selected Regions | ADNI AD-CN $\Delta\alpha_f$ | | | | |
|---|---|---|---|---|---|
| | Surface Area | Cortical Thickness | GWR | Curvature | Sulcal Depth |
| Inferior Temporal Gyrus | -0.43 | 1.26 | 1.98 | 2.19 | -0.10 |
| Parahippocampal Gyrus | -0.69 | 1.00 | 1.77 | 1.54 | 0.04 |
| Superior Occipital Gyrus | -1.00 | 0.17 | 0.45 | 0.54 | -0.07 |
| Angular Gyrus | -1.03 | 0.28 | 1.15 | 1.13 | -0.10 |
| Pars Opercularis | -0.63 | 0.21 | 0.81 | 1.29 | -0.21 |
| **Selected Networks** | | | | | |
| Default Mode Network | -0.53 | 0.54 | 1.32 | 1.53 | -0.14 |
| Limbic/Paralimbic System | -0.50 | 0.90 | 1.52 | 1.71 | -0.08 |
| Somatomotor Network | -0.54 | 0.38 | 1.17 | 0.94 | -0.12 |

**Table 3. Selected ADNI AD-CN $\Delta\alpha_f$ values in years.** Representative ADNI AD-CN $\Delta\alpha_f$ values indicating the relative change in LBAGs for a given ablated feature across locations and cohorts. A positive $\Delta\alpha_f$ indicates that the feature contributes more strongly to disease-specific aging, while a negative $\Delta\alpha_f$ indicates that the feature contributes more strongly to normal aging. Values near zero indicate that the feature affects both groups similarly. Regions and networks were selected based on within-cohort outlier status and biological priors.

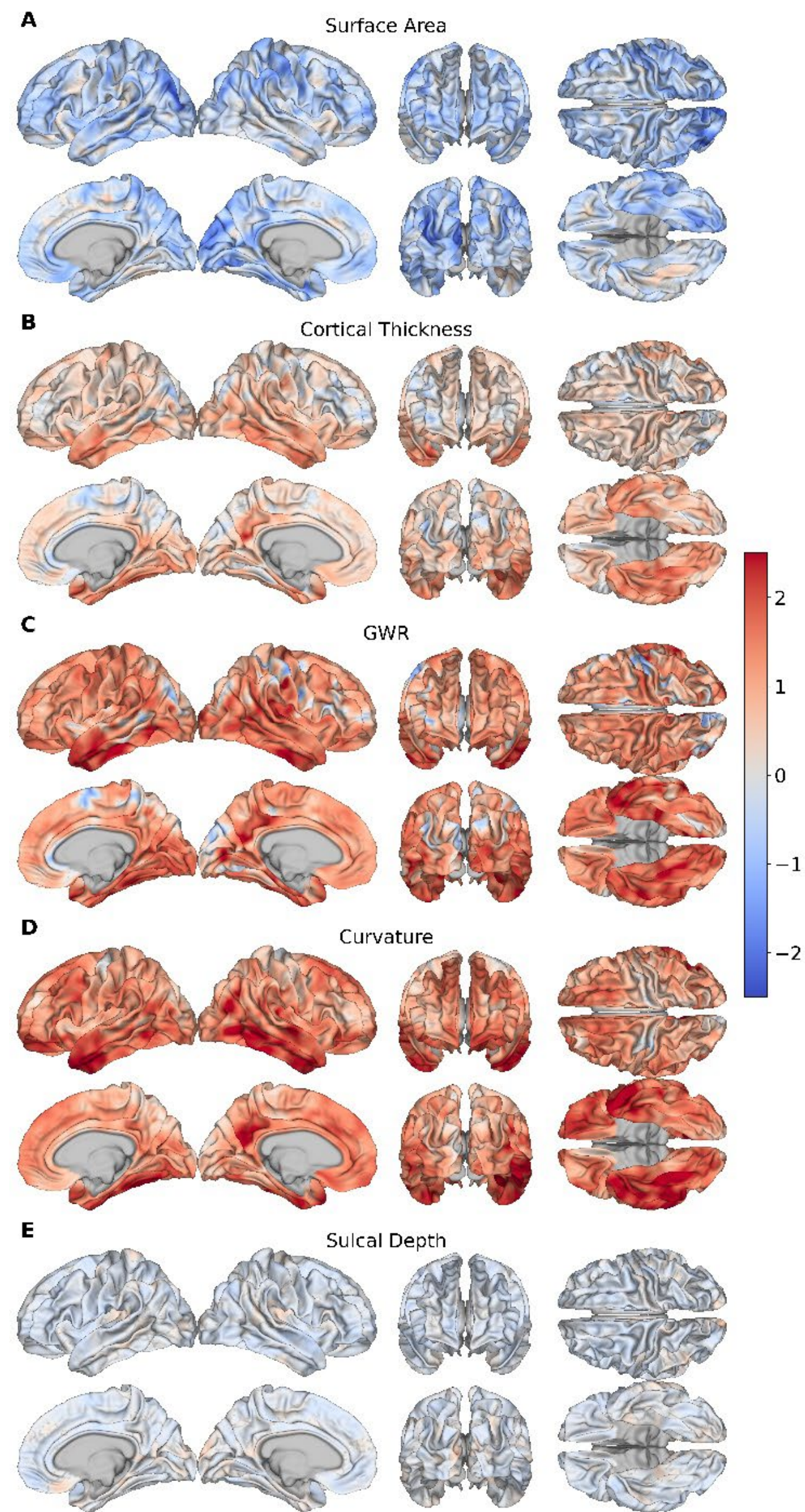

**Figure 3. $\Delta\alpha_f$ for ADNI AD-CN subjects.** Group-level $\Delta\alpha_f$ maps of the relative influence of each cortical feature in driving AD-specific aging. (A) Surface area, (B) cortical thickness, (C) GWR, (D) curvature, (E) sulcal depth. Units are shown in years.

**Regressing LBAGs against cognition.** To determine if changes in LBAGs accompany changes in cognition, we regressed LBAGs against cognitive scores for the AD and CN cohorts. We chose the inferior temporal gyrus and postcentral gyrus as regions of interest because these regions had the largest and smallest bilateral AD ΔLBAGs respectively. Across both regressions, no significant associations were observed between LBAGs and cognitive scores for the CN cohort. By contrast, several cognitive tests were associated with regional LBAGs for the AD cohort, surviving Benjamini-Hochberg correction. For the inferior temporal gyrus (**Figure 4**), these included the Montreal Cognitive Assessment (MoCA; $\beta = 1.41$), Alzheimer's Disease Assessment Scale-11 (ADAS11; $\beta = 1.25$), Rey Auditory-Verbal Learning Test (RAVLT) Immediate Recall ($\beta = 1.23$), Mini-Mental State Examination (MMSE; $\beta = 1.19$), Functional Activities Questionnaire (FAQ; $\beta = 1.01$), Clinical Dementia Rating Sum of Boxes (CDRSB; $\beta = 1.00$), Digit Symbol Substitution (DSST; $\beta = 0.96$), RAVLT Percent Forgetting ($\beta = 0.67$), and RAVLT Learning ($\beta = 0.51$) tests. The postcentral gyrus had a similar pattern (**Figure 4**), but demonstrated smaller $\beta$ coefficients which failed to reach significance for $DSST$ ($p = 0.20$) or RAVLT Percent Forgetting ($p = 0.06$). The postcentral gyrus also maintained significant associations with MoCA ($\beta = 0.70$), ADAS11 ($\beta = 0.70$), RAVLT Immediate Recall ($\beta = 0.87$), MMSE ($\beta = 0.71$), FAQ ($\beta = 0.74$), CDRSB ($\beta = 0.75$), and RAVLT Learning ($\beta = 0.57$) performance. Regression statistics are available in **Table S5**.

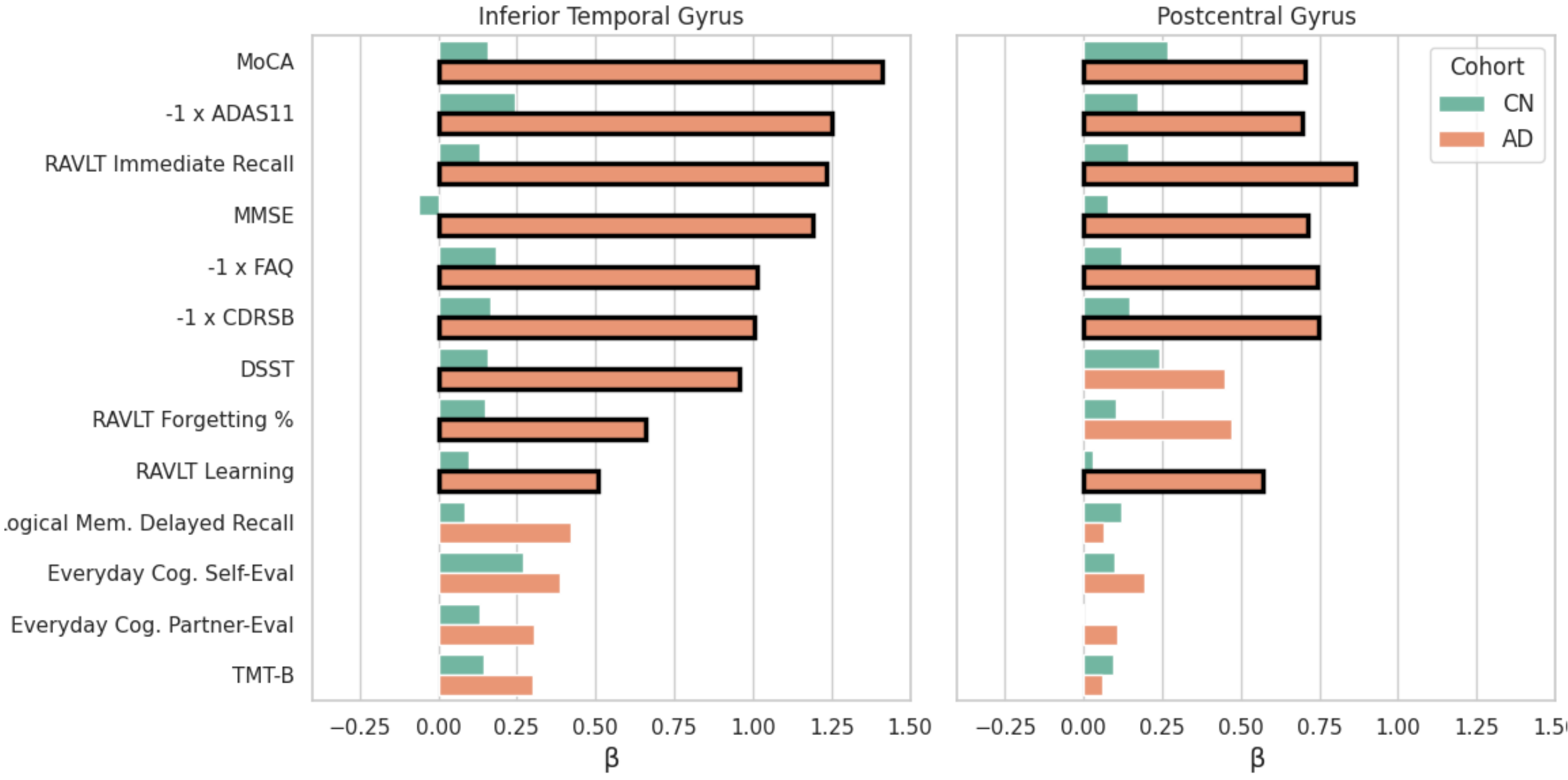

**Figure 4. Regression results between LBAGs and cognitive scores.** LBAG was regressed against CA, cognitive score, sex, and years of education. Cognitive scores were z-score standardized and transformed (indicated by -1 x) such that a positive β coefficient indicates worsening performance with increasing LBAGs for each test. Bars show β coefficients for each test/cohort pair, using regional LBAGs from the inferior temporal gyrus and postcentral gyrus, respectively. Green bars indicate regressions for the CN cohort, while orange bars indicate regressions for AD subjects. Bars in bold indicate significant associations (adjusted $p < 0.05$).

**Sex Differences.** To identify differences in aging across sexes, we computed ΔLBAGs across female and male subjects in the ADNI CN cohort (**Figure S2**). Female subjects had younger LBAGs at every region across the cortex, with the largest effects being in sulci. These included the precentral sulcus (superior = -0.97 y, inferior = -0.88 y), postcentral sulcus (-0.93 y), and superior frontal sulcus (-0.88 y). All LBAGs and ΔLBAGs are reported by region and network in **Tables S6-7**.

**Local comparison to LBA-SotA.** To benchmark our model against CNN-based approaches to LBA, we compared the spatial distribution of AD ΔLBAGs across the cortical surface in our model to the LBA-SotA (Figure 5). Accordingly, these comparisons are intended to assess relative sensitivity to AD-specific aging patterns rather than establish identical architectural benchmarking conditions. The LBA-SotA had only minor variation across lobes when comparing CN and AD subjects. By contrast, while our GNN model identified a comparable degree of aging in parietal, frontal, insular, and occipital lobes, a pronounced increase in temporal lobe aging was observed in ADs.

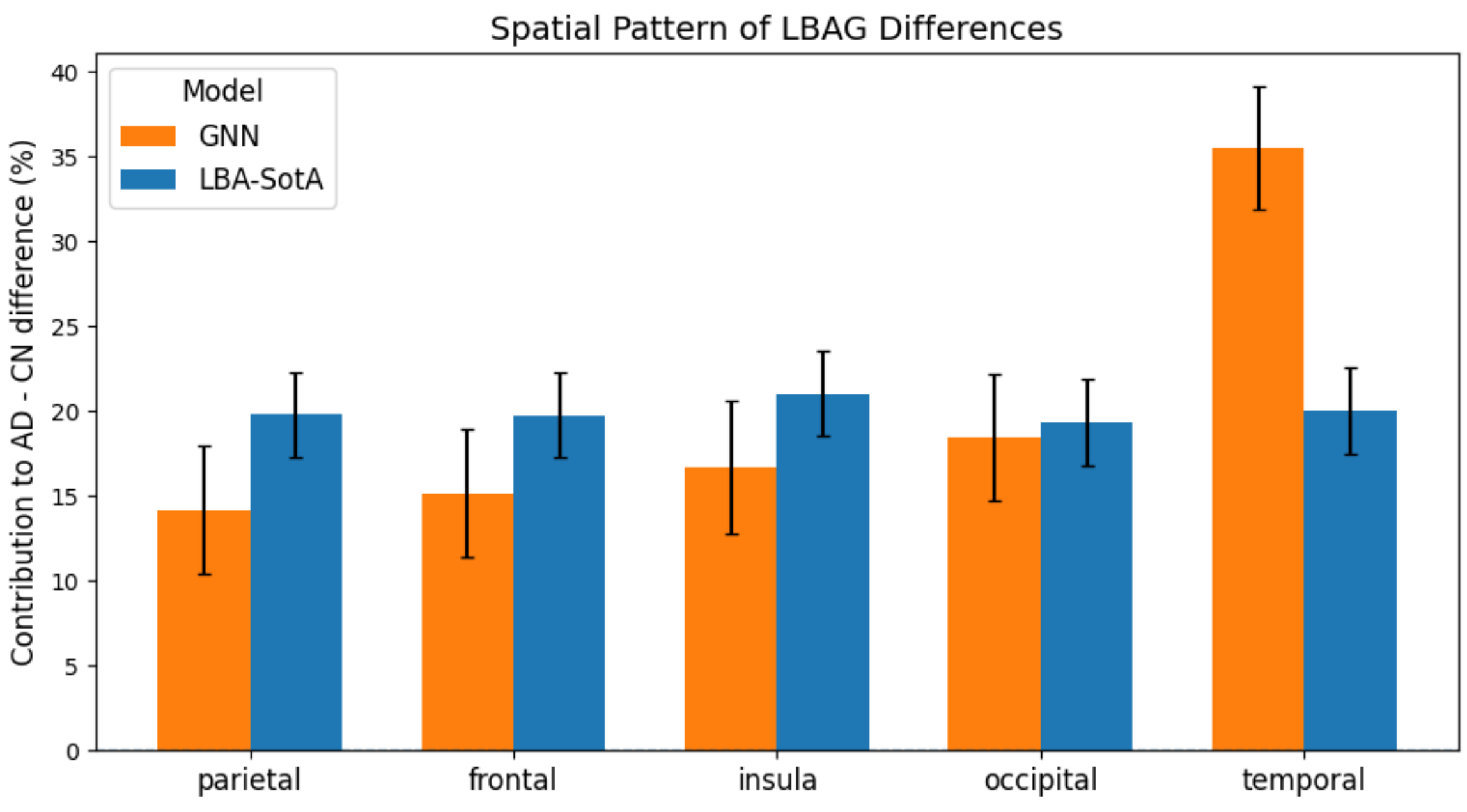

**Figure 5. AD ΔLBAGs by lobe in the GNN and LBA-SotA.** Values represent within-model normalized AD ΔLBAGs by lobe and model. Evaluation subjects were matched between models to ensure fair comparison. Error bars denote the standard error of the mean. ΔLBAGs were normalized within each model such that absolute values sum to 100%, indicating the relative contribution of each lobe to the total disease effect. Orange bars represent the GNN, while blue bars represent the LBA-SotA.

## Discussion

**Technical novelty.** Our LBA model addresses the limitations of GBA by modeling local aging patterns directly on the cortical surface, enabling direct local comparisons at high spatial resolution. Existing LBA methods leverage CNNs applied to volumetric data, where local neighborhoods are defined by sliding fixed cubic kernels through the brain [35-37]. This assumption of Euclidean geometry disregards the folded topology and structural irregularity of the cortical surface, risking spatial distortions that conflate anatomically distant regions [41-44]. By representing the cortical surface in its native non-Euclidean topology, the model learns directly from a more faithful representation of the geometric structure of the cortex, potentially preserving clinically meaningful biological signals [72-74]. Our resampling strategy further reinforces these benefits by preserving anatomical structure across all stages of model processing. Collectively, these design choices support the detection of subtle, subject specific local aging patterns that may be lost in alternative approaches.

**Model performance and comparison to previous work.** To assess model performance relative to the LBA-SotA [37], we compared the MAE and AD ΔLBAGs across both models. The GNN model achieved a 0.31-year lower MAE than the LBA-SotA and revealed markedly different patterns of pathological aging under comparable evaluation conditions. The GNN model indicated substantially greater aging in the temporal lobes of AD subjects, whereas the LBA-SotA identified approximately uniform LBAG differences across all lobes. In concordance with previous work, our model identifies greater relative aging in the temporal lobe as a characteristic pathological feature of AD [75-78]. Given that the primary novelty of an LBA model is its local sensitivity, this finding suggests that our GNN model may be better suited for LBA estimation than CNN alternatives, consistent with both its theoretical advantages and lower MAE.

**Interpretation of structural findings.** Aging in CN individuals exhibited the largest LBAGs throughout the visual, parietal, and prefrontal association cortices, with notable negative LBAGs in the LIM. Parietal and prefrontal association cortices are among the earliest sites of morphometric aging, and although the visual association cortex is relatively resistant to atrophy in middle-aged adults, notable degeneration is a well-established finding in older adults [79-81], consistent with the older age distribution of the ADNI CN cohort. Conversely, the LIM is robust against aging effects across the lifespan, especially in GM [79, 82, 83].

In subjects with MCI, the parahippocampal gyrus exhibited one of the largest bilateral ΔLBAGs, consistent with prior literature establishing this region as central to AD pathology [3, 75, 78]. These effects were relatively diffuse compared to AD subjects, aligning with the heterogeneity of MCI phenotypes as a transitional stage between normal and AD aging [84, 85].

In AD subjects, the inferior temporal gyrus and parahippocampal gyrus exhibited some of the largest increases in brain aging, highlighting these regions as sensitive markers of disease progression [86-88]. The inferior temporal gyrus functions as a central hub in AD, facilitating the global propagation of tau across the cortex and undergoing severe and early degeneration that correlates with cognitive decline [88-91]. At the network level, we observed large positive ΔLBAGs across the DMN and LIM, with only weak effects in the SMN. The DMN and LIM are strongly associated with AD-related neurodegeneration, whereas the SMN is relatively spared until later stages of the disease [75-77, 92-95]. Notably, the DMN, LIM, and parahippocampal gyrus all converge within medial temporal regions, which often serve as the locus of AD pathology [78]. Because AD pathology is thought to propagate preferentially through the DMN and LIM [75, 76, 96], these findings may reflect intrinsic network vulnerabilities and distributed pathological involvement through topologically connected cortical systems.

The persistent involvement of the parahippocampal gyrus across MCI and AD supports its role as a sensitive marker of disease progression, with the propagation of AD-associated aging through biological networks highlighting the importance of preserving cortical topology when modeling pathological signals. Cumulatively, these findings suggest a progression from association cortex–dominated aging in healthy individuals, to globally distributed alterations in MCI with transient peaks of AD-associated aging, and finally to medial temporal and network-specific degeneration in AD.

**Feature ablation.** Surface area $\Delta\alpha_f$ values were negative across most of the cortex, with the strongest effects in the parietal and visual association cortices. This suggests that regional surface area variability is more reflective of normative aging than AD pathology. Consistent with this interpretation, prior work has shown that although surface area changes across the lifespan, it remains relatively preserved in AD [81, 97, 98].

By contrast, $\Delta\alpha_f$ values for curvature, GWR, and cortical thickness were elevated in the inferior temporal gyrus, parahippocampal gyrus, LIM, and DMN, mirroring the spatial distribution of AD ΔLBAGs. These findings align with evidence implicating cortical thickness [75, 81, 98], GWR [94, 99, 100], and curvature [13, 101-103] in AD-related neurodegeneration. Among these features, curvature and GWR exhibited the strongest effects, consistent with prior findings that these features capture early, microstructural changes that precede overt atrophy [94, 100, 101, 104]. Cortical thickness, on the other hand, was distinguished by its disproportionately larger LIM signal relative to the DMN. Because the LIM exhibits relative preservation of cortical thickness throughout healthy aging, this effect may reflect a selective breakdown of this preservation, whereas the DMN thins significantly in both healthy aging and AD [82, 92, 93]. Sulcal depth ablation produced weak, near-zero $\alpha_f^c$ values for both CN and AD cohorts, suggesting little explanatory contribution—potentially due to redundancy with other morphometric features [105, 106].

Surface area preferentially indexes normative aging trajectories, especially in parietal and visual association cortices, whereas curvature and GWR are more sensitive to AD-specific alterations. Cortical thickness demonstrated milder pathological specificity overall, though its effects peaked disproportionately in the LIM relative to other networks. Sulcal depth provided little explanatory power when other features were present. Collectively, these findings clarify how distinct morphometric features capture complementary aspects of brain aging.

**Relationship to cognition.** Across regions, the inferior temporal gyrus exhibited larger effect sizes than those observed in the precentral gyrus, and demonstrated additional significant associations with RAVLT Percent Forgetting and DSST scores in the AD cohort. Neither region exhibited significant interactions between CN LBAGs and cognitive scores, reflecting the clinical utility of these tests in AD diagnosis rather than in CN populations [107-111]. The stronger interactions between inferior temporal gyrus LBAGs and cognitive tests over precentral gyrus LBAGs are consistent with the differential vulnerability of these regions to AD pathology. The inferior temporal gyrus directly influences the global severity of AD through tau propagation [88-91], whereas the precentral gyrus is relatively spared until later stages of the disease [75]. Accordingly, LBAG variability in the inferior temporal gyrus would be expected to better track global disease burden, and by extension broader cognitive deficits, confirming that LBAGs reflect measurable clinical decline beyond structural epiphenomena.

**Sex differences.** Our sex analysis indicated that females were younger than males across the entire cortex, with the largest differences being in sulci. These findings align with the prevailing view that female brains appear younger than male brains [112, 113], especially in sulci [114-116]. This demonstrates the sensitivity of our model to sex-related differences in brain atrophy.

**Limitations of study design.** A limitation of our study lies in the composition of the training and testing sets. The training set was skewed toward older adults and not fully age-matched, which may have reduced model generalizability. It was also sex-imbalanced (female-to-male ratio of 1.3:1) and, in the case of NACC, included multiple scans from the same participants to minimize the over-representation of UKBB in our training set. Another limitation pertains to the testing set, which only included older adults. This is appropriate for MCI- and AD-focused analyses, but precludes evaluation of model performance in younger populations.

A more theoretical limitation of our approach stems from the geometric structure of the cortical surface, and how we leverage this to propagate information. From a graph-theoretic standpoint, the cortical mesh suffers from poor global connectivity, creating bottlenecks that impede the flow of information. This structural rigidity contributes to over-squashing, a well-documented phenomenon that hinders the model's capacity to model long-range dependencies [117-119]. Various solutions to this problem have been proposed—most notably graph rewiring—but these come with significant trade-offs. Crucially, rewiring distorts the graph's native topology, complicating anatomic interpretability. In addition, many of these techniques risk introducing over-smoothing, hindering local variability and potentially obscuring local brain-aging patterns [119-121]. In our own experiments, existing rewiring methods either failed to yield performance gains [122, 123], or were computationally infeasible due to the mesh's scale [119, 124]. Future studies may benefit from the implementation of models that combine accurate local representations with more explicit multiscale organization. This could be achieved through hierarchical graph structures that facilitate information flow between fine- and coarse-scale descriptions, or by enriching graph representations to better capture long-range or cross-scale interactions beyond fixed local neighborhoods.

## Conclusions

We introduced a GNN for LBA estimation that leverages morphometric features to produce biologically grounded maps of cortical aging. Our model achieves lower MAE than the LBA-SotA while identifying AD-specific aging patterns that better align with established neuroanatomical vulnerability profiles. The model identified association cortices as particularly vulnerable in normative aging, with the parahippocampal gyrus implicated in both MCI and AD, highlighting its potential as an early marker of disease progression. AD subjects demonstrated large LBAGs in medial temporal regions, the DMN, and the LIM, consistent with the preferential involvement of medial temporal regions and large-scale cortical networks in AD-related decline. Feature-level analyses further suggest that different morphometric properties capture distinct aspects of brain aging, with surface area more sensitive to normal aging than to AD-related decline, whereas curvature and GWR demonstrate heightened sensitivity to AD pathology. Regressing LBAGs on cognitive scores recapitulated these findings, grounding our structural analyses in clinical measures of impairment. Collectively, these findings introduce and validate a framework for directly modeling cortical aging patterns as a function of morphology, demonstrating the potential of vertex-level brain age estimation as a means of quantifying local aging with respect to disease, morphology, and cognition.

**Acknowledgments**

S.D.A. is thankful to Rachel Fox for her assistance in debugging. A.I. is grateful to Profs. Dag Aarsland and Richard Siow for hosting him at the Institute of Psychiatry, Psychology & Neuroscience of King's College London, where part of this research was conducted during a sabbatical leave from the University of Southern California.

**Sources of Funding**

This study received support from the US National Institutes of Health (NIH) under grant R01 AG 079957, from the Hanson-Thorell Research Scholarship Fund, and from anonymous donors. Data collection and sharing for this project was funded by the Alzheimer's Disease Neuroimaging Initiative (ADNI) through NIH grant U01 AG 024904 and DoD ADNI through DoD award number W81-XWH-1220012. ADNI is funded by the National Institute on Aging, the National Institute of Biomedical Imaging and Bioengineering, and through generous contributions from the following: AbbVie, Alzheimer's Association; Alzheimer's Drug Discovery Foundation; Araclon Biotech; BioClinica, Inc.; Biogen; Bristol-Myers Squibb Company; CereSpir, Inc.; Cogstate; Eisai Inc.; Elan Pharmaceuticals, Inc.; Eli Lilly and Company; EuroImmun; F. Hoffmann-La Roche Ltd and its affiliated company Genentech, Inc.; Fujirebio; GE Healthcare; IXICO Ltd.; Janssen Alzheimer Immunotherapy Research & Development, LLC.; Johnson & Johnson Pharmaceutical Research & Development LLC.; Lumosity; Lundbeck; Merck & Co., Inc.; Meso Scale Diagnostics, LLC.; NeuroRx Research; Neurotrack Technologies; Novartis Pharmaceuticals Corporation; Pfizer Inc.; Piramal Imaging; Servier; Takeda Pharmaceutical Company; and Transition Therapeutics. The Canadian Institutes of Health Research is providing funds to support ADNI clinical sites in Canada. Private sector contributions are facilitated by the Foundation for the NIH (www.fnih.org). The ADNI grantee organization is the Northern California Institute for Research and Education, and the study is coordinated by the Alzheimer's Therapeutic Research Institute at the University of Southern California. ADNI data are disseminated by the Laboratory for Neuro Imaging at the University of Southern California. Research reported in this publication was also supported by the National Institute on Aging of the NIH under award U01 AG 052564.

This research has been conducted using the UK Biobank Resource, under application number 47656. The views, opinions, and/or findings contained in this article are those of the authors and should not be interpreted as representing the official views or policies, either expressed or implied by the NIH or any other entity acknowledged here. The funding sources had no role in study design; in the collection, analysis and interpretation of data; in the writing of the report; and in the decision to submit the article for publication. Computational resources were made available using a generous gift from an anonymous donor family. Portions of this article were edited using ChatGPT and Claude to improve readability and language. All intellectual content and conclusions are the authors' own.

## Disclosures

The authors have no competing interests to declare.

## Supplementary Figures

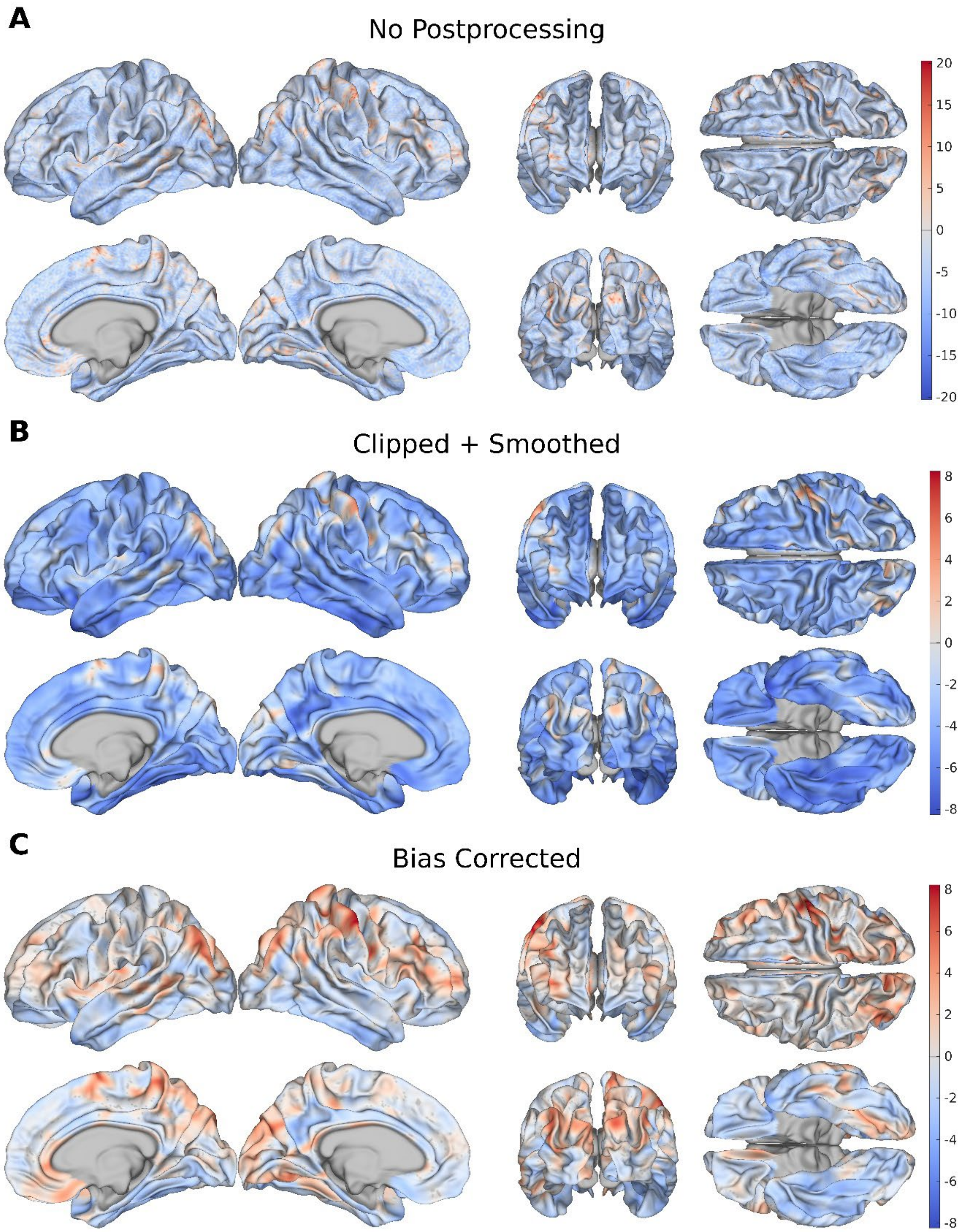

**Figure S1. LBAGs at different processing stages.** (A) LBAGs before any processing. (B) LBAGs after clipping and smoothing. (C) LBAGs after bias correction.

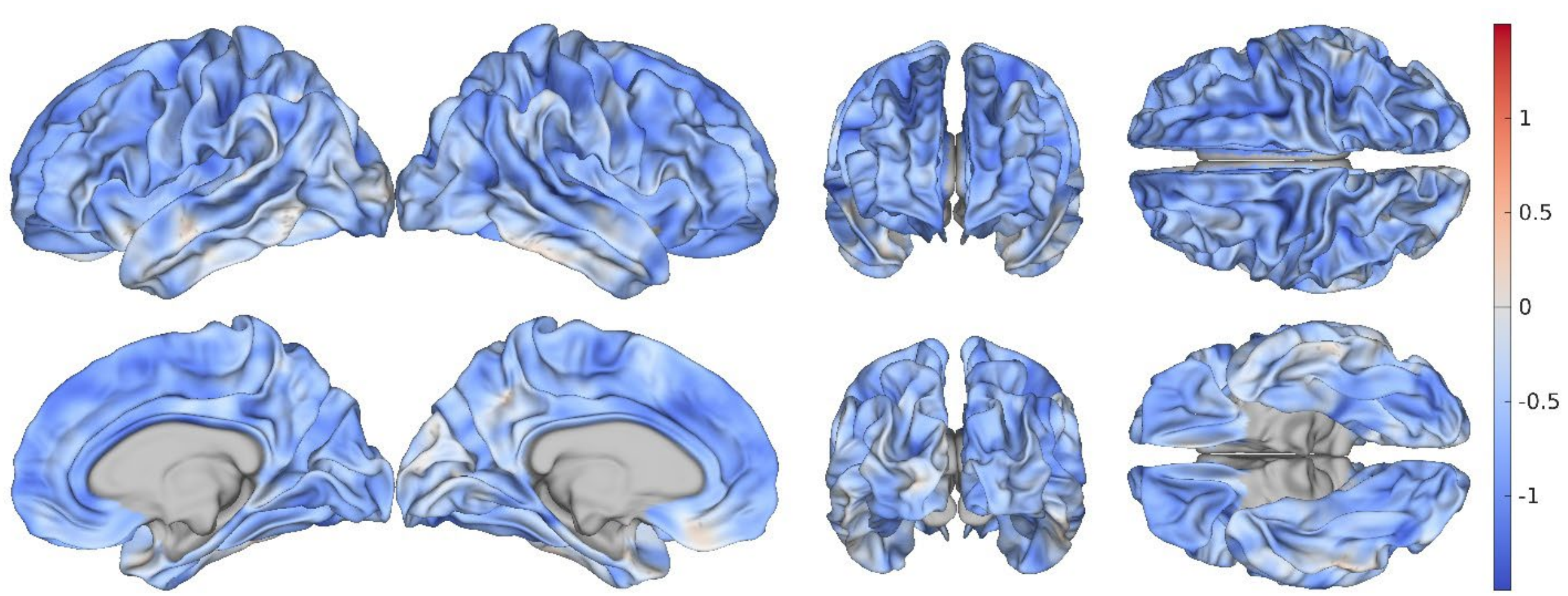

**Figure S2. Bias-corrected ΔLBAGs for sex.** Vertex-wise ΔLBAGs between Female and Male CNs (Female – Male).